# Mathematical Foundations for the Economy of Giving

*W.P. Weijland*

Informatics Institute, Faculty of Science, University of Amsterdam, Science Park 904, 1098 XH Amsterdam, the Netherlands

**Abstract.** This paper shows how we can build a model for transactions when goods are given away in the expectation of a later settlement. In settings where people keep track of their social accounts we are able to redefine concepts like account balance, yield curve and the law of diminishing returns. The model provides us with a result that expresses how people have a structural preference for one recipient over the other regardless the actual account balance. Hence a building block in the social fabric of a community. Finally, a fundamental theorem is presented to show how suppliers and recipients use their account balance in order to reach an equilibrium in the exchange of goods much like the traditional balance between supply and demand.

## I. INTRODUCTION

In monetary economy Léon Walras (1834-1910)[1] and Alfred Marshall (1842-1924)[2] are often cited as founders of the general equilibrium theory in economics and among the first economists to measure and calculate equilibria at the intersection of the classic supply and demand curves. Western economic modelling is unthinkable without the views and tools that they proposed at the turn of the 18th and 19th century.

In a world without money of account one cannot use such supply and demand curves to find optimum economic transactions in the market place. In this world goods are given away or are being exchanged for other goods without any currency to express the value of such transaction. Such a world occurs more often than one may think. At times of disaster or extreme poverty when monetary systems are failing for example. And in family systems and rural village communities relationships can be more important than money of exchange and even replace this money as such. The modern Internet also facilitates communities where information and/or goods are shared or given away for free without any monetary concept behind it.

In this article we develop a mathematical model for such economies by means of a system of social credit instead serving as an alternative money of account. Every individual keeps track of a mental bank account of social credit in relation to every other person individually. If you give away something to someone else you will not take it as a loss. Instead you will receive credits to be settled at some later point. Both individuals keep track of these credits and constantly take a (subjective) view on their current balance.

---

[1] Éléments d'économie politique pure, Léon Walras, 1874.
[2] Principles of Economics, Alfred Marshall, 1890.

After having defined the basic elements of social credit accounts and yield functions we will examine the consequences of the Law of Diminishing Returns[3]. Next we will introduce the Highest Yield Rule to decide on the preference for each individual in a situation of choice. Finally we analyze how a moneyless market eventually comes to an equilibrium similar to the ones found by Walras and Marshall.

## II     CREDIT: THE ACCOUNT BALANCE

In order to formulate our key results we need some concepts and terminology first.

DEFINITION II.1      A *multiset* is a non-ordered, final set with multiple occurrences of elements. For example, we write M = [1,1,2,5,7,4,7,7,8] for a nine-element multiset of natural numbers 1, 2, 4, 5, 7 and 8 with occurrences 2, 1, 1, 1, 3, 1 respectively[4]. A *multiset over a set A* is a multiset containing only (multiple) elements from set A.

If M is a multiset over A then all domain elements in A not occurring in M are said to have occurrence zero in M. We write a∈M if the occurrence of a in M is greater than zero.

A multiset K is a *multi-subset* of multiset L, notation: K⊆L, if all elements occur at most as many times in K as they do in L. A *multipair* is a multiset with just two elements.

For multisets K and L we write K ∪ L for the multiset containing only elements from K or L and in which all elements occur precisely as many times as the sum of their occurrences in K and L respectively.

DEFINITION II.2      An *entity* is a person, group, village, tribe, people or any other subject able to interact with other entities through the transfer of goods, where a *good* is any type of product, service or favor with a (positive) economic value. All entities under consideration together are referred to as the *community*.

DEFINITION II.3      The act of offering a good *a* to all entities by entity P is called the *supply of a by P*. Notation: P —$^a$→. Entity P is called a *supplier* of *a*.
The act of accepting a good *a* from any entity by Q is called a *demand of a by Q*, written as: —$^a$→Q. Entity Q is called a *recipient* of *a*.

The supply P —$^a$→ and demand —$^a$→Q together may lead to a *transaction of a from P to Q*, notation: $t$ = P —$^a$→Q. Such transaction *t* consists of a multipair [P —$^a$→, —$^a$→Q] of the supply and a demand of the same product *a*. A transaction is said to be a *transaction between P and Q* if it is a transaction from P to Q or a transaction from from Q to P. P and Q are *involved* in transaction *t*.

DEFINITION II.4      The *Supply-Demand Space* SDS is defined as:
SDS = { P —$^a$→, —$^b$→Q : for all entities P, Q and all goods *a*, *b*}.
A *state* is a multi-subset over SDS. Transaction $t$ = P —$^a$→Q is called *admissible* in state S if P —$^a$→ ∈S and —$^a$→Q ∈S.

---

[3] Originally developed as a concept by Ricardo (1772-1823).
[4] More precisely, we note that a multiset is a set (the universe) together with a mapping from its elements to $\mathbf{N_0}$ (respective occurences).

A multiset of transactions T = [$t_1$, $t_2$, $t_3$, ..., $t_k$] is *admissible* in state S if $\cup_{i \in \mathbb{N}} t_i \subseteq S$.
In other words: S caters for all supply and demand needed to enable all transactions in T.

DEFINITION II.5    The *standard model* consists of a sequence of pairs ($S_i$, $T_i$)$_{i \in \mathbb{N}}$ of states $S_1$, $S_2$, $S_3$, ... and multisets of transactions $T_1$, $T_2$, $T_3$, ... such that each $T_i$ is admissible in $S_i$.

Where $S_i$ is the space of all offerings and acceptances at point i, $T_i$ can be viewed as the set of transactions actually realized at that point.

DEFINITION II.6    For each entity P and at each point i we assume the existence of a *yield function* $\cdot^P$ mapping each possible transaction $t$ from $T_i$ to a real number such that:

(i) If P is not involved in $t$ then $t^P = 0$
(ii)    If $t = P \xrightarrow{a} Q$ for some Q and $a$ then $t^P \geq 0$
(iii)   If $t = Q \xrightarrow{a} P$ for some Q and $a$ then $t^P \leq 0$.

One may look at $t^P$ as the value attached by P to a certain transaction. Note that we have such a yield function in each point i so that P may value the same transaction differently in different points.

The yield functions of P and Q are *competible* if: $t^P + t^Q = 0$, for all transactions $t$ between P and Q. This implies that P and Q share the same view on the value of each transaction that they enter into. The notion of competibility also allows us to define a relative account balance between P and Q:

DEFINITION II.7    At any point i the *account balance* $A_{i,P,Q}$ of P versus Q is defined as:

(i) $A_{0,P,Q} = 0$
(ii)    $A_{i+1,P,Q} = A_{i,P,Q} + \sum_t t^P$, where $t$ ranges over all transactions in $T_i$.

Note that if the yield function is competible then $A_{i,P,Q} + A_{i,Q,P} = 0$ voor all i.

## III.    GIVING: THE LAW OF DIMINISHING RETURNS

### YIELD CURVES

Let us assume that the transaction $t = P \xrightarrow{a} Q$ means that product $a$ is actually given away to Q by P. P could look at such transaction as an investment in his relation with Q, hoping that Q will return a certain value in the future. Thus the yield $t^P$ of a transaction $t$ can be seen as the *present value of a future settlement of transaction t as perceived by P*.

In reality one may expect such expectation of a future settlement to depend on how much indebted Q already is with respect to P: if Q is deeply in the red with P, than P may value any next transaction $t$ less than when P is in the red with Q. This leads to the following assumpion:

ASSUMPTION III.1    We assume that the yield $t^P$ is a function of the account balance $A_{i,P,Q}$ between P and Q. As the yield $t^P$ does not depend on the state index i, we write $A_{P,Q}$ to refer to the account balance between P and Q regardless the state we are at.

So the we also have:

DEFINITION III.2    Given entities P and Q and transaction $t = P \xrightarrow{a} Q$ the *yield curve of P* is the function that for every account balance $A_{P,Q}$ produces the yield $t^P$.

DEFINITION III.3    The *Law of Diminishing Returns* expresses that for all entities P the yield curve is monotonically decreasing.

The Law of Diminishing Returns is a plausible assumption when one realises that the higher the account balance $A_{P,Q}$ the lower the yield of transaction $t$: the more positive $A_{P,Q}$, the more P is 'in the black' with Q, hence the more Q is 'in the red' with P, and the less likely it is (in the eyes of P) that Q will ever settle transaction $t$ that is added as yet another debt on Q's behalf.

In graphical form the Law of Dimishing Returns may look as follows:

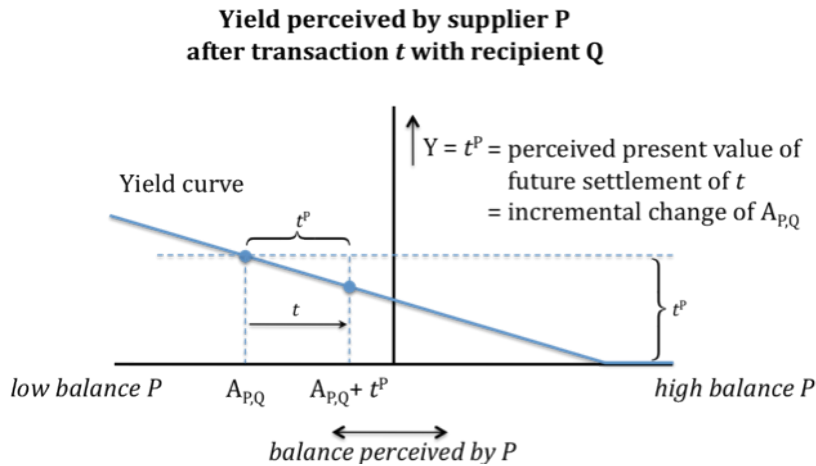

*Graph III.1*

In this graph we have assumed the yield curve to be linear and decreasing until it reaches Y = 0, after which it remains flat. Obviously, when the yield of transaction $t$ is zero there is no economic reason for P to pursue transaction $t$ with Q any longer. Notice that the dotted lines together enclose a square shape representing the step $t$ down the curve.

Repeating transaction $t$ leads to the step diagram in Graph III.2 below:

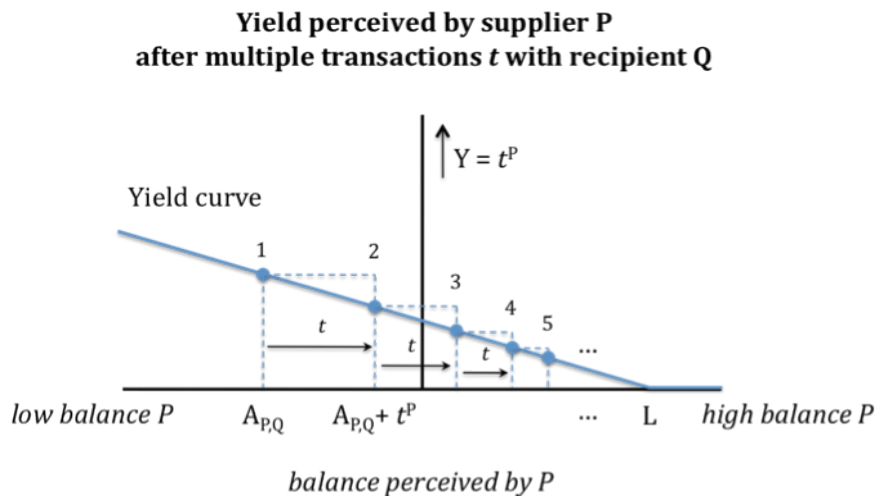

*Graph III.2*

Intuitively this graph says that if P provides an endless supply of good *a* to Q without any redemption whatsoever then the yield of such transaction gradually decreases to zero. When the yield curve is linear the yield approaches zero asymptotically and the balance between P and Q approaches limit L. The points on the graph are called *valuation points* (pairs of account balance and yield). These valuation points follow the curve downwards expressing a stepwise decrease of the yield of *t*. Notice that the dotted lines form a sequence of squares decreasing in size.

In algebraic terms we can describe the yield curve by the following equation:

$$y = -a.x + b, \text{ for certain } 0 \leq a < 1, 0 \leq b$$

where y is the yield of *t* and x is the balance $A_{P,Q}$ perceived by P. If we assume that the points 1, 2, 3, … on Graph III.2 are written as $(x_1,y_1), (x_2,y_2), (x_3,y_3),…$ then it follows easily from the construction of the graph that:

$$x_{k+1} = x_k + y_k$$

and so:

$$y_{k+1} - y_k = (-a.x_{k+1} + b) - y_k = -a.(x_k + y_k) + b - y_k = -a.x_k - a.y_k + b - y_k =$$
$$= -a.x_k + b - (a+1).y_k = y_k - (a+1).y_k = -a.y_k$$

so we find:

(1)     $y_{k+1} = (1-a).y_k$      or equivalently:     $y_k = (1-a)^k.y_1$.

Parameter a is a called the *yield coefficient*[5]. The higher a the steeper the curve and the faster the loss of faith in a future settlement of *t*. The term (1 – a) can be seen as a stepwise discount factor to the initial yield $y_1$.

Parameter b is called the *nominal value of t*: if the balance between P and Q is neutral, hence $A_{P,Q} = 0$, the yield of *t* is precisely b. Throughout this paper we assume the nominal value to be non-negative.

The repeated transaction in this section fits into the standard model of section II as follows. For all i∈**N** choose $S_i := [P \xrightarrow{a}, \xrightarrow{a} Q]$. Then $T_i := [P \xrightarrow{a} Q]$ are admissible for all i. The sequence $(S_i, T_i)_{i \in \mathbf{N}}$ represents the infinite offering of *a* by P to the always accepting recipient Q.

### MULTIPLE RECIPIENTS

Suppose entity P has goods *a* and *b* to give away. Suppose entity Q accepts *a* and R accepts *b*. Which transaction will P choose? In our model we assume P will choose the transaction that generates the *highest yield*.

DEFINITION III.4     A *choice axiom* is a rule that for every state S selects one out of all multisets of transactions that are admissible in S.

DEFINITION III.5     For every entity P the *Highest Yield Rule* $HYR^P$ is the choice axiom that selects the admissible multiset T for which $\sum_{t \in T} t^P$ is maximal. The axiom that $HYR^P$ applies to all P is simply written as HYR.

To see how the HYR works look at Graph III.3 below. Assume at every point P may choose between two transactions: $t = P \xrightarrow{a} Q$ and $t' = P \xrightarrow{b} R$. Initially, P perceives his account balance with Q at $A_{P,Q}$ and with R at $A_{P,R}$. Point 2 and 1 are the corresponding valuation points on the yield curves with Q and R respectively.

---

[5] For now we assume a<1. The situation that a≥1 implies that the yield curve reaches zero in one step. It requires further research to interpret the meaning of such situation in theory.

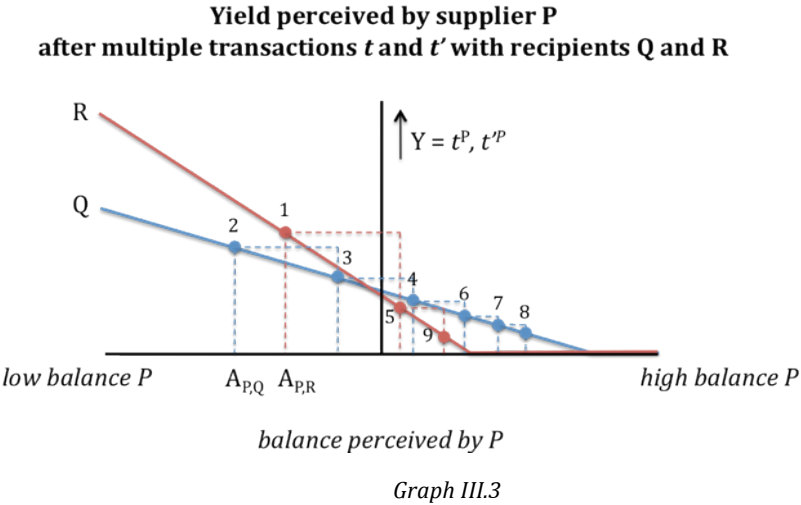

*Graph III.3*

Clearly, the HYR selects valuation points top down. It starts by selecting point 1 and the corresponding transaction $t'$. After this valuation point 1 moves down to point 5 as the account balance between P and R shifts to the right and transaction $t'$ is now valued at a much lower level. In the next step HYR prefers point 2 and the corresponding transaction $t$ since the yield of $t$ at point 2 is much higher than the yield of $t'$ at point 5. One can easily see that this results in an infinite process in which both $t$ and $t'$ are selected infinitely many times.

ASSUMPTION III.6   (pairwise independence) In the above analysis we have assumed that the yield curves themselves remain unaffected in every step. A transaction $t$ from P to Q does not affect P's yield curve with R and vice versa.

P does not select the recipient as such, but he selects the transaction that produces the highest increase in account balance regardless with whom, Q or R.

For the case in this section the state space $S_i$ as in the standard model of section II can be defined as follows. For all i∈**N** choose $S_i := [P \xrightarrow{a}, \xrightarrow{a} Q, \xrightarrow{a} R]$. Now the question is: which admissible sequence $(T_i)_{i \in \mathbf{N}}$ does respect the highest yield rule (HYR) in every point i? It turns out that in the first nine steps $T_i$ looks as follows:
$T_i := [P \xrightarrow{a} R, P \xrightarrow{a} Q, P \xrightarrow{a} Q, P \xrightarrow{a} Q, P \xrightarrow{a} R, P \xrightarrow{a} Q, P \xrightarrow{a} Q,$
$P \xrightarrow{a} Q, P \xrightarrow{a} R, ...]$. The sequence $(S_i, T_i)_{i \in \mathbf{N}}$ represents the infinite offering of $a$ by P to the always accepting recipients Q and R.

**ULTIMATE DISTRIBUTION THEOREM**
In the infinite process of selecting $t$ and $t'$ in Graph III.3 both $t$ and $t'$ are selected infinitely many times. However, they are not necessarily selected equally often. In the graph one can already see that in the first nine steps of the process the valuation points on the curve for Q outnumber the ones on the curve for R. Thus Q gets more than a 'fair share' of the transactions offered by P.

The yield curves for Q and R can be described by the following equations:

Q:   $y = -a \cdot x + b$, $0 \leq a < 1$, $0 \leq b$
R:   $z = -c \cdot x + d$, $0 \leq c < 1$, $0 \leq d$.

We write $y_k$ for the consecutive valuation points on the curve for Q and $z_i$ for those on the curve for R. From formula (1) in this section (see above) we learn:

$y_k = (1 - a)^k \cdot y_1$.
$z_i = (1 - c)^i \cdot z_1$.

The point at which the HYR switches from the one curve to the other is where $y_k = z_i$ at which point P has selected *t* (recipient Q) already k times and *t'* (recipient R) precisely i times. So the ratio of selecting *t* over *t'* is equal to k/i. Then we find:

$y_k = z_i$  hence:   $(1 - a)^k \cdot y_1 = (1 - c)^i \cdot z_1$

Assuming $y_1 > 0$, $z_1 > 0$ we find:

$k \cdot \log(1 - a) + \log y_1 = i \cdot \log(1 - c) + \log z_1$

and:

$k/i = \log(1 - c)/\log(1 - a) + (\log z_1 - \log y_1)/(i \cdot \log(1 - a))$

As i gets bigger the last term approaches to zero. Hence:

(2)   $k/i \approx \log(1 - c)/\log(1 - a)$,   as $k, i \to \infty$.

DEFINITION III.7   If $0 \leq a < 1$ is the yield coefficient of the linear yield curve of transaction $t = P \xrightarrow{a} Q$ from P to Q, then the *ultimate credit ratio* (UCR) is defined as:  $C_Q := -1/\log(1 - a)$.

THEOREM III.8 (Ultimate Distribution Theorem)      Assume for all $i \in \mathbb{N}$ we have $S_i := [P \xrightarrow{a}, \xrightarrow{a} Q, \xrightarrow{a} R]$. The yield curve of (P, Q, *t*) has UCR = $C_Q$ and the yield curve of (P, R, *t'*) has UCR = $C_R$. If initially we have $y_1 > 0$, $z_1 > 0$, then the relative number of times that the HYR selects *t* is equal to $C_Q/(C_Q + C_R)$.

PROOF     With (2) we have: $k/i = C_Q/C_R$ and so the relative number of times *t* is selected in the process is $k/(k+i) = C_Q/(C_Q + C_R)$, as $k, i \to \infty$. □

The fractions $C_Q/(C_Q + C_R)$ and $C_R/(C_Q + C_R)$ represent the *ultimate distribution* between Q and P, or more precisely: between *t* and *t'*. The UCR can be seen as a 'preference factor' ('like factor') with P of Q and R respectively. Notice that the UCR does not depend on the nominal value b or d. Neither does it depend on the initial account balance $A_{P,Q}$ or $A_{P,R}$. It solely depends on the gradient of the yield curve.

**IV.     TRADE:  STRIVING FOR BALANCE**

In this chapter we will utilize the tools developed in the previous sections to model a market in which suppliers are recipients and vice versa. We will refer to such a market as a *trading market*. We will look for equilibriums in supply and demand.

**TRADING WITH TWO ENTITIES**

Suppose we have two entities P and Q that consititute the entire trading market. P is supplier of good *a* of which Q is the recipient. Q is supplier of good *b* of which P is the recipient.

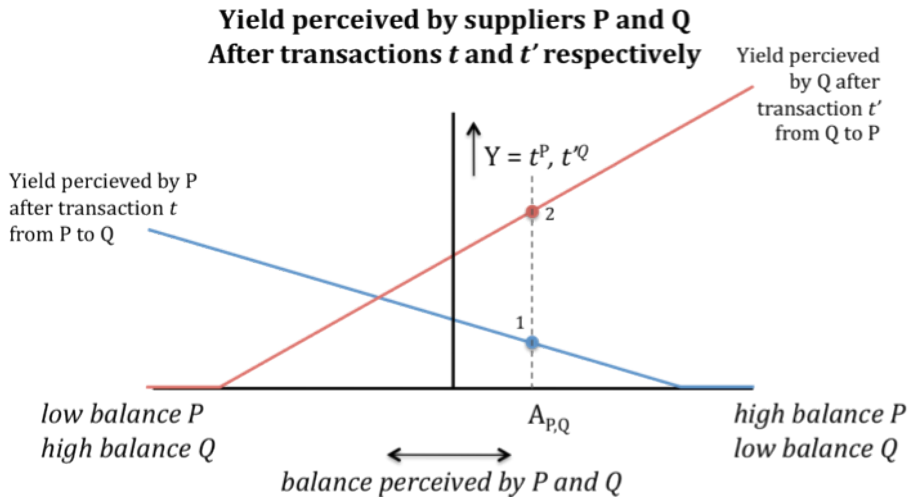

*Graph IV.1*

In Graph IV.1 we have displayed the situation. Notice that we have drawn the yield curve of Q in the reverse direction: from left to right Q's balance runs from high to low, whereas P's balance runs from low to high. This is to make sure that at each transaction the balance accounts move into the same direction.

From now on we assume that P and Q have the same perception as to their account balance:

ASSUMPTION IV.1    Yield functions are mutually competible: $A_{P,Q} + A_{Q,P} = 0$ for all P, Q.

Given balance $A_{P,Q}$ in Graph IV.1 it is obvious that the yield of transaction $t'$ at point 2 on Q's curve is higher than that of transaction $t$ at point 1 on P's curve. The question is: how will this affect the trading process between P and Q? We will explore two cases: one with simultaneous availability of goods and one with alternating availability of goods.

**SIMULTANEOUS AVAILABILITY OF GOODS**

The first case is one where $a$ and $b$ will be simply traded simultaneously as represented by $S_i := [P \xrightarrow{a}, Q \xrightarrow{b}, \xrightarrow{a} Q, \xrightarrow{b} P]$. Then $T_i = [P \xrightarrow{a} Q, Q \xrightarrow{b} P]$ is admissible at each instance and the yield diagram looks as in Graph IV.2:

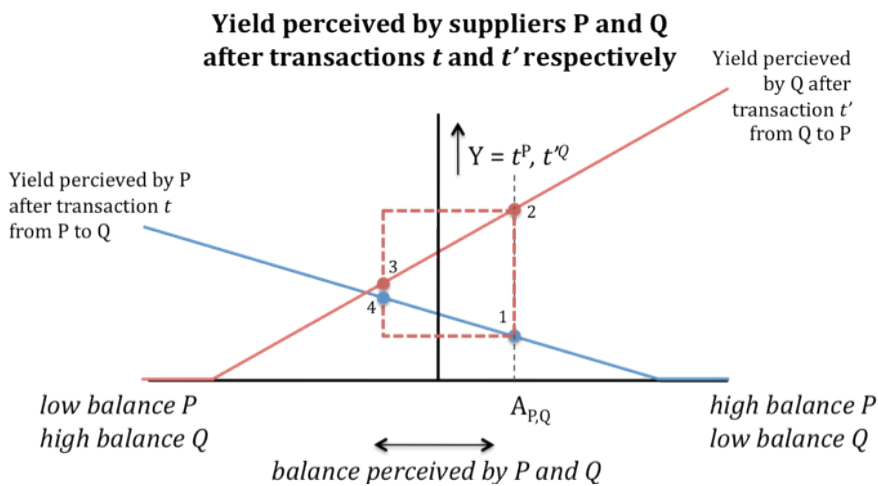

*Graph IV.2*

This can be understood as follows: at point $A_{P,Q}$ the transaction $t = P \xrightarrow{a} Q$ is valued at valuation point 1 whereas the transaction $t' = Q \xrightarrow{b} P$ is valued much higher at point 2. When the two transactions are performed simultaneously the resulting shift in account balance is precisely the difference between the two yields and $A_{P,Q}$ moves to the left in favour of the balance of Q.

Now the process repeats itself be it with a much smaller difference in valuation between points 3 and 4. It is not difficult to see that this process converges to the intersection point of the two yield curves where there valuations are equal. Hence the equal supply of goods *a* and *b* will ultimately lead to an equilibrium where both goods are valued equally.

**ALTERNATING AVAILABILITY OF GOODS**

The question is whether we will also find such equilibriums when *a* and *b* are not traded simultaneously but rather one by one. Such as for example in the case where we have:
$S_{2i-1} := [Q \xrightarrow{b}, \xrightarrow{a} Q, \xrightarrow{b} P]$ and $S_{2i} := [P \xrightarrow{a}, \xrightarrow{a} Q, \xrightarrow{b} P]$ for all $i \in \mathbb{N}$. Then $T_{2i-1} = [Q \xrightarrow{b} P]$ and $T_{2i} = [P \xrightarrow{a} Q]$ are admissible for all i. They are alternating transactions.

The yield diagram this time looks different. We obtain an iterative process pictured in Graph IV.3 below:

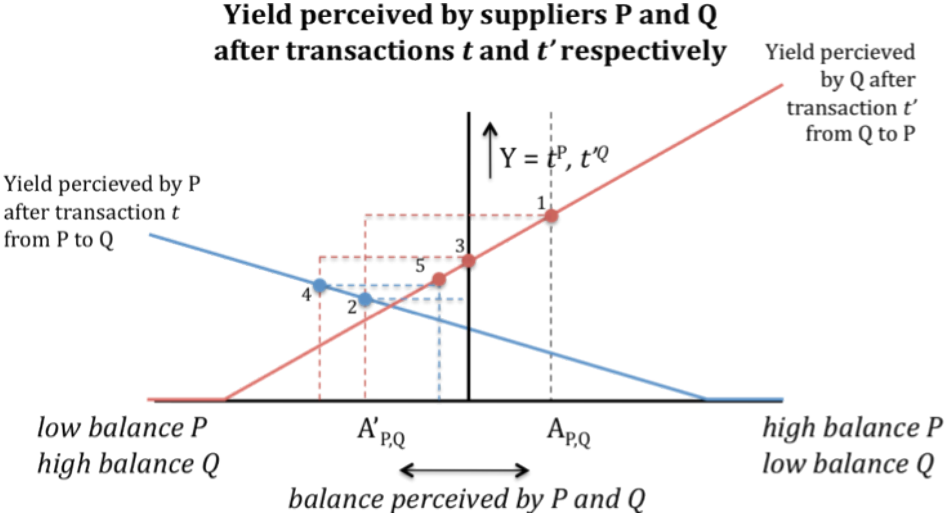

*Graph IV.3*

At point 1 the account balance is $A_{P,Q}$ and $t' = Q \xrightarrow{b} P$ is the only admissible transaction (as in every odd point). So, the account balance moves to $A'_{P,Q}$. At $A'_{P,Q}$ the only admissible transaction is $t = P \xrightarrow{a} Q$ valued at point 2. The account balance moves back to under valuation point 3. And so on. The account balance moves back and forth alternately. The question is: will this iterative process settle eventually? The answer is yes as we will see in the following paragraphs.

**THE CANONICAL EQUILIBRIUM**

Let us take the yield curves from Graph IV.3. Suppose we can draw a square with the bottom on the *x*-axis, and the upper corners on the curves of P and Q respectively. See also Graph IV.4 where the square has vertices at $x_\infty$, $x'_\infty$, $y_\infty$ and $z_\infty$.

Now assume the account balance between P and Q is at $x_\infty$ and that $t = P \xrightarrow{a} Q$ is the only admissible transaction. Following the square shape we end up in $x'_\infty$ where the only admissible transaction is $t' = Q \xrightarrow{b} P$. From $x'_\infty$, however, we return to $x_\infty$ through transaction $t'$, exactly where we were before. We find an equilibrium consisting of two valuation points $y_\infty$ and $z_\infty$ and the iteration will endlessly alternate between them.

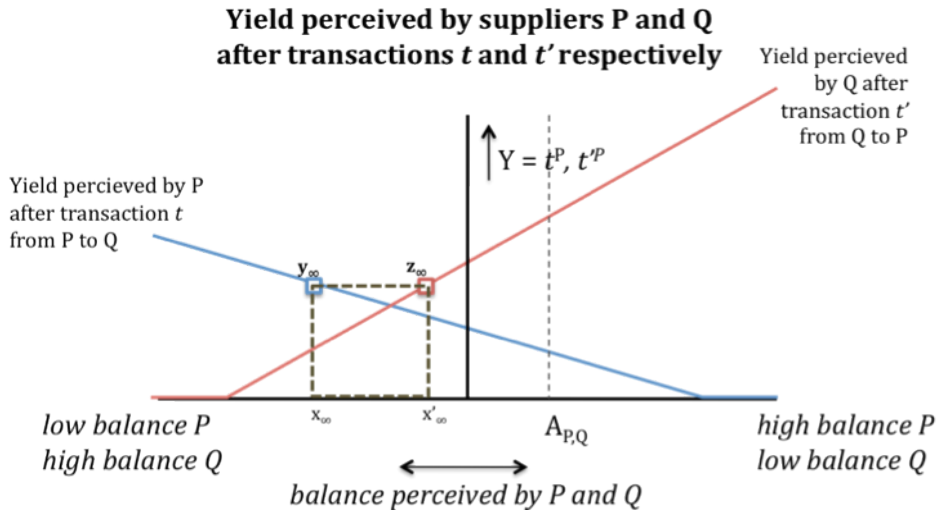

*Graph IV.4*

THEOREM IV.2     If $P(x)$ and $Q(x)$ are linear yield curves with yield coefficients $0 < a, c < 1$ then there exists exactly one square with the bottom on the *x*-axis, and the upper corners on the curves of $P(x)$ and $Q(x)$ respectively.

PROOF  Like in section III we can describe the two linear yield curves as follows:

    P:     $y = -a.x + b$    $0 \le a < 1, 0 \le b$
    Q:     $z = c.x + d$    $0 \le c < 1, 0 \le d$.

Notice that the yield coefficient of Q has an opposite sign because we put Q's curve in reverse direction. The intersection $(x_s, y_s) = (x_s, z_s)$ of these curves can be calculated at:

(1)     $x_s = (b - d) / (a + c)$
(2)     $y_s = z_s = (ad + bc) / (a + c)$.

This point is not an equilibrium since an iteration starting from $(x_s, y_s)$ will not return to $(x_s, y_s)$ again (except for the trivial case $x_s = y_s = z_s = 0$). The equilibrium $x_\infty$, $x'_\infty$, $y_\infty$ and $z_\infty$ from Graph IV.4 can be calculated from the following equation:

     $y_\infty = x'_\infty - x_\infty = z_\infty$

implying:

     $y_\infty = -a.x_\infty + b = x'_\infty - x_\infty$    so:      $x'_\infty = (1 - a).x_\infty + b$
     $z_\infty = c.x'_\infty + d = x'_\infty - x_\infty$    hence:    $x_\infty = (1 - c).x'_\infty - d$

By substitution we get:

     $x'_\infty = (1 - a).x_\infty + b = (1 - a).(1 - c).x'_\infty - (1 - a).d + b$ and so:
     $(a + c - ac).x'_\infty = b - (1 - a).d$.

Since $0 < a, c < 1$ we have $(a + c - ac) > 0$, and so:

(3)     $x'_\infty = (b - (1 - a).d) / (a + c - ac)$
(4)     $x_\infty = ((1 - c).b - d) / (a + c - ac)$
(5)     $y_\infty = (ad + bc) / (a + c - ac) = z_\infty$.

From these three equalities we find that the square $x_\infty$, $x'_\infty$, $y_\infty$ and $z_\infty$ from Graph IV.4 exists and there is only one such square. □

DEFINITION IV.3    The equilibrium from Theorem IV.2 is called the *canonical equilibrium of P and Q* since it always exists, provided yield coefficients <1.

In Graph IV.4 the canonical equilibrium implies a negative account balance for P in both points $x_\infty$, $x'_\infty$. This can be explained as follows: the nominal value of *t'* is higher than that of *t*: transaction *t'* has more 'weight' than *t*. This excess in weight can only be compensated by 'inflation', i.e.: by a high balance of Q and a low balance of P. At such a point P and Q are able to come to an arrangement of alternate exchange of *t* and *t'*.

**THE CONVERGENCE THEOREM**

THEOREM IV.3 (Convergence Theorem)    If $P(x)$ and $Q(x)$ are intersecting linear yield curves with yield coefficients $0<a,c<1$ then the iteration process of Graph IV.3 converges to the canonical equilibrium regardless the account balance $A_{P,Q}$ at the starting point.

PROOF  Graph IV.3 suggests that the sequence $x_1, x_3, x_5,...$ and $x_2, x_4, x_6,...$ both converge. Suppose they do then it is clear that both convergence points together form an equilibrium and since the canonical equilibrium is the only existing equilibrium it is evident that the limits are $x'_\infty$ and $x_\infty$ respectively. We will prove that $x_i - x_{i+2}$ converges to zero exponentially, thereby proving the theorem.

Like in section III we can describe the two linear yield curves as follows:

P:    $y = -a.x + b$    $0 \leq a < 1$, $0 \leq b$
Q:    $z = c.x + d$    $0 \leq c < 1$, $0 \leq d$.

Again, notice that the yield coefficient of Q has an opposite sign because we put Q's curve in reverse direction. For odd numbers i we have:

$x_i - x_{i+2} = x_i - (x_{i+1} + t^P) = x_i - (x_{i+1} - a.x_{i+1} + b) = x_i - ((1-a).x_{i+1} + b) =$
$= x_i - ((1-a).(x_i - (c.x_i + d)) + b) = x_i - ((1-a).((1-c).x_i - d) + b) =$
$= x_i - (1-a).(1-c).x_i + (1-a).d - b = (a + c - ac).x_i + (1-a).d - b$.

Hence also:

$x_{i+2} - x_{i+4} = (a + c - ac).x_{i+2} + (1-a).d - b$

And therefore:

$(x_i - x_{i+2}) - (x_{i+2} - x_{i+4}) = (a + c - ac).(x_i - x_{i+2})$, so:

and:

$(x_{i+2} - x_{i+4}) = (1 - a - c + ac).(x_i - x_{i+2}) =$
$= (1-a).(1-c).(x_i - x_{i+2})$.

Since $0<a,c<1$ we have that $(1-a).(1-c)$ is positive and smaller than 1 and we find that $(x_i - x_{i+2})$ converges to zero exponentially.

For even numbers i we find:  $(x_{i+2} - x_{i+4}) = (1-c).(1-a).(x_i - x_{i+2})$ and following the same logic $(x_i - x_{i+2})$ converges to zero exponentially. This concludes our proof. □

The convergence theorem is relevant because it shows that in our model whenever goods *a* and *b* are available in equal supply they will eventually be traded as equal in value.

A variant of this case can be specified as: $S_i := [P \xrightarrow{a} , Q \xrightarrow{b} , \xrightarrow{a} Q, \xrightarrow{b} P]$ for all $i \in \mathbb{N}$. Then in each state $S_i$ both $P \xrightarrow{a} Q$ and $Q \xrightarrow{b} P$ are admissible and we can use the HYR as a choice axiom to select one. One can prove that also this time the convergence theorem holds. In both cases the standard models are alternating in the long run. However, the model using the HYR may select the same transaction finitely many times in a row before permanently alternating towards the canonical equilibrium.

## V. FURTHER RESEARCH

### THE CONCEPT OF GIVING

The concept of giving does not seem to have been studied in the literature within a mathematical framework comparable to the one in this paper. However, a similar concept, namely that of *promises*, has been studied in a context of logic by Bergstra and Burgess [2]. Interestingly, a 'promise' may be conceived reciprocal to a 'gift' in the economical sense: if P values the gift of *a* to Q at the present value of a future settlement by Q, then one may argue that the acceptance of *a* by Q is equivalent to the promise of Q to eventually settle the debt resulting from such gift.

Besides the economical approach much has been written about 'giving' in the context of altruism, philantropy and many other viewpoints. Even when we eliminate egocentric giving – aimed at creating social credit or just a 'warm glow' – it has been proven that still a substantial amount of altruism remains [3], which is supported by many other studies in the literature on philantropy [4]. So the proposed model for the Economy of Giving as described in this paper does by no means cover all there is to say about giving in general. However, we have not been able to identify any reference in the literature, modelling the economic aspects of giving in quantifiable terms as we did here.

### PERISHABLE AND NON-PERISHABLE GOODS

In the standard model we have proposed states $S_i$ to represent the space of all available supply and demand at point i. A question is whether a transaction of a certain good affects the supply or demand at any later point. A *perfectly perishable good* ceases to exist in the transition to the next state (they may be consumed or destroyed). If all goods are perfectly perishable then we can do without the notion of *stock*. Likewise, a *perfectly non-perishable* good is a good that cannot 'disappear'. At each state it will be supplied by its owner or be kept in stock (which requires a separate notion in the model).

The extension of the standard model with stock may be simple. It is an open question how the ultimate distribution theorem and the convergence theorem are affected after having done so.

### THE HIGHEST YIELD RULE (HYR)

The HYR maximizes the yield over all possible transactions in $S_i$ in favour of suppliers. This is plausible in a *supply market* where recipients outnumber - or at least match - suppliers and hence suppliers dictate market prices (in traditional economy this is called a 'buyer's market'). The cases dealt with in sections III and IV are all supply markets.

However, in case we have multiple suppliers and only one recipient the scene changes drastically. For example, choose $S_i := [P^1 \xrightarrow{a}, P^2 \xrightarrow{a}, \xrightarrow{a} Q]$ where two suppliers 'compete' for one recipient[6]. The Highest Yield Rule now flips to the negative since Q now has a choice whether it takes *a* from $P^1$ or from $P^2$. One may expect that Q wishes to *minimize* the cost of *a*, or – more formally – minimize the present value of the future settlement of the receipt of good *a*. We could assume recipients Q to have a *cost curve* for every supplier P and every potential transaction $P \xrightarrow{a} Q$. How this works out in theory is still open.

#### THE ULTIMATE DISTRIBUTION THEOREM GENERALIZED

One can easily extend the ultimate distribution theorem of section III from the case with one supplier P and two recipients to the one with one supplier and multiple recipients $Q^1, \ldots, Q^N$ with corresponding transactions $t^i = P \xrightarrow{a} Q^i$ for $1 \le i \le N$. The state space $S_i$ can be written as $S_i := [P \xrightarrow{a}, \xrightarrow{a} Q^1, \ldots, \xrightarrow{a} Q^N]$. Suppose the recipients $Q^1, \ldots, Q^N$ have UCR's $C^1, \ldots, C^N$ calculated from there respective yield coefficients with P. One easily finds that eventually the relative number of times that $t^k$ (with recipient $Q^k$) is selected by the HYR is equal to:

(1)     $C^k / (C^1 + \ldots + C^N)$.

We conjecture that this result can be generalized further to one for the model with state space $S_i := [P^1 \xrightarrow{a}, \ldots, P^M \xrightarrow{a}, \xrightarrow{a} Q^1, \ldots, \xrightarrow{a} Q^N]$ for $M \le N$, with one product *a* and with M suppliers and N recipients where no supplier can be a recipient or vice versa. The condition $M \le N$ makes the model a supply market so that we do not run into the complication of reformulating the highest yield rule. Clearly, all P's are served at all times and the recipients are selected by the M highest yields each time.

#### RECURRENT STATE SPACES

In most of the models so far we have taken $S_i$ constant over time (i.e. over the integer i). It looks like many of the results in this paper can be reproduced when the sequence $(S_i)_{i \in \mathbf{N}}$ is *recurrent* or *rational* (like rational numbers).

DEFINITION V.1                 A sequence $(S_i)_{i \in \mathbf{N}}$ is *recurrent* if there exist indices $k, m \in \mathbf{N}$ such that for all $i > m$: $S_i = S_{i+k}$. The smallest of such numbers k is the *dimension* of the recurrence of $(S_i)_{i \in \mathbf{N}}$.

It looks as if the ultimate distribution theorem can be generalized to recurrent state sequences. If the recurrent sequence has dimension k then one can look at it as if we are running k models in parallel, each one of which produces its own ultimate distribution. The overall ultimate distribution in the composite model can be found by taking the average over all distributions of the individual models.

#### THE CONVERGENCE THEOREM GENERALIZED

In section IV we started out with the example of simultaneous availability of goods. We showed how this leads to a one-point equilibrium where the respective valuations of the goods *a* and *b* are equal. Notice that the chosen model $(S_i)_{i \in \mathbf{N}}$ for simultaneous availability of goods has dimension 1 (see definition V.1).

---

[6] Note that competition among suppliers is plausible in the case of perfectly perishable goods. Without giving it away the good perishes and a potential value is lost.

In the second example we looked at a model of dimension 2, with two alternating states and the canonical equilibrium consisting of two points between which the system eventually alternates. These valuation points are with equal valuation expressing that in the canonical equilibrium *a* and *b* are valued equally.

Now let us look at the case of alternating availability of goods where product *a* is offered twice as much as product *b*. For example, for all i∈**N**:

$S_{3i-2} := [Q \xrightarrow{b}, \xrightarrow{a} Q, \xrightarrow{b} P]$
$S_{3i-1} := [P \xrightarrow{a}, \xrightarrow{a} Q, \xrightarrow{b} P]$
$S_{3i} := [P \xrightarrow{a}, \xrightarrow{a} Q, \xrightarrow{b} P]$.

Then $T_{3i-2} = [Q \xrightarrow{b} P]$ is admissible at each point 1, 4, 7, 10,... and $T_{3i-1} = T_{3i} = [P \xrightarrow{a} Q]$ is admissible at all other points i. In words: every transaction $Q \xrightarrow{b} P$ is followed by two consecutive transactions $P \xrightarrow{a} Q$. It turns out that this system leads to an equilibrium consisting of three points that are visited alternately. Further, in that equilibrium the yields of two transactions $P \xrightarrow{a} Q$ add up to the yield of $Q \xrightarrow{b} P$. This expresses that in the canonical equilibrium the average transaction value of *b* is two times the average transaction value of *a*, precisely in accordance to their relative supply.

Our conjecture is that we may generalize the convergence theorem such that:

1. Any model with:
    a. a finite state dimension, say: k
    b. at least as many suppliers as recipients for all states and all products
    c. with only yield coefficients <1
    d. respecting the HYR
   has a canonical equilibrium which consists of at most k valuation points.

2. In the canonical equilibrium the average price of a good relative to the price of another good is equal to the average supply of that good relative to the supply of that other good.

This may lead to a general theorem that in a supply market the relative average price of a good is determined only by its relative supply.

##### NON-LINEAR YIELD CURVES

The question may rise how the theory presented here depends on the assumption that yield curves are linear. Our conjecture is that the ultimate distribution theorem and the convergence theorem hold for any model where the yield curves are monotonically decreasing with yield coefficient <1.

A potential outline of proof may consist of the following: at any point a non-linear graph moves in the direction of the tangent which itself is linear. So at that point the curve moves towards an ultimate distribution and/or canonical equilibrium determined by the tangent's linear parameters. Since all yield curves are monotonically decreasing with coefficient <1 the standard model sequence operates as a contraction in the space of ultimate distributions and canonical equilibriums. Hence the sequence settles at a certain fixed point.

##### AN ALGEBRAIC APPROACH

One can specify the standard models in terms of recursive equations as in process algebra [1]. The equation P = *a* stands for P offering *a*. The equation P = *a*.P stands for repeatedly offering *a*. Similarly, the equation Q = *a̲*.Q stand for Q repeatedly accepting *a*.

The operator (*a*|*a̱*) represents the transaction of a from P to Q. This way it may be possible to establish a one-to-one correspondence between finite algebraic specifications and recurrent state sequences.

### GAME THEORY

The link between our model and that of *game theory* is quite straightforward. One can imagine Q deciding to not pursue the acceptance of *a* from P because it would disqualify him for the receipt of some other good that he values more. If one drops the assumption that all entities be pairwise independent it is possible to allow each entity to determine a strategy that optimizes the yield over the whole model.

### APPLICATIONS IN COMPUTER SCIENCE

Applications of the model in this paper are not limited to the domain of economics alone. In computer science we find systems consisting of loosely coupled networks that cooperate by sharing memory and processing power. The Economy of Giving can be used as a protocol allowing computers to make excess capacity available to other computers in the network while optimizing the likelihood that such favor will ever be returned.